# Evaluating the Impact of Urban Road Topology on Quantum Approximate Optimization: A Comparative Study of Planned and Organic Networks in Islamabad and Karachi(Pakistan)

Shumaila Ashfaq[1], Abdul Sami Rao[2], Roha Ghazanfar Khan[3]

Software Engineering Department, NED University, Karachi, Pakistan

Corresponding author: shumailaashfaq@cloud.neduet.edu.pk

## Abstract

The performance of shallow-depth quantum optimization algorithms is known to depend strongly on problem structure, yet the role of real-world network topology remains poorly understood. In this work, we study how urban graph structure influences the behaviour of the Quantum Approximate Optimization Algorithm (QAOA) at depth $p = 1$. Using street-network subgraphs extracted from two cities in Pakistan with contrasting urban designs—a planned city (Islamabad) and an organically grown city (Lyari)—we analyze probability concentration, approximation quality, and performance variability on the minimum vertex cover problem. By comparing classical brute-force solutions with QAOA outcomes, we show that planned topologies yield more reliable convergence, while organic networks exhibit higher variance and a greater tendency toward trivial solutions. Our results suggest that urban structure primarily affects the robustness rather than the average quality of shallow QAOA solutions, highlighting the importance of higher-order structural heterogeneity in shaping low-depth quantum optimization landscapes. This research is vital because it bridges the gap between abstract quantum theory and the chaotic reality of our physical world, proving that the way we build our cities directly impacts our ability to optimize them. By identifying how "topological DNA" influences algorithmic success, this work enables the development of more resilient quantum solutions for critical infrastructure, such as smart power grids and emergency response routing. Ultimately, these insights benefit society by paving the way for more efficient, data-driven urban management that can reduce resource waste and improve the quality of life in both planned and organically growing metropolitan areas.

## Keywords

# 1. Introduction

As modern infrastructure networks grow in scale and complexity, classical optimization methods increasingly struggle to deliver efficient solutions [1]. This challenge raises a critical question of whether quantum algorithms, such as the Quantum Approximate Optimization Algorithm (QAOA) [2], can fundamentally reshape how we optimize real-world networks [3]. Modern cities, supply chains, and communication systems are built on complex spatial networks [4] whose underlying structure strongly influences the efficiency of resource and information flow [4]. As these systems expand, traditional combinatorial methods face significant computational bottlenecks, motivating a transition toward quantum computing as a potential paradigm shift for solving large-scale combinatorial problems [5].

In the realm of classical optimization, critical tasks such as routing and resource allocation are often formulated as combinatorial tasks. Many of these, including Max-Cut [6] and the Minimum Vertex Cover (MVC) [7], are classified as NP-hard, rendering exact solutions computationally infeasible as the problem size scales. To address these challenges on Noisy Intermediate-Scale Quantum (NISQ) devices, the QAOA has emerged as a prominent hybrid quantum–classical heuristic [8]. The algorithm operates by encoding an optimization task into a cost Hamiltonian—frequently using Ising or Quadratic Unconstrained Binary Optimization (QUBO) formulations [9]—and iteratively evolving a quantum state through parameterized circuits guided by classical optimizers [10].

Despite the academic interest in QAOA, a significant portion of current research relies on abstract benchmarks or synthetic graph generators [11]. Consequently, there is a limited understanding of how real-world geographical topologies, such as urban street networks, impact the algorithm's performance. The intricate relationship between structural graph properties, including node degree distribution [12], algebraic connectivity [13], and local clustering [14], and the resulting quality and stability of QAOA solutions remains largely underexplored [11]. This represents a critical gap in the literature, as optimization problems in transportation and logistics are inherently constrained by physical geography and spatial order [15].

The primary objective of this paper is to analyze how different geographical network topologies affect the performance of quantum optimization, specifically focusing on the Minimum Vertex Cover problem. By comparing QAOA outcomes against classical baselines across various network sizes and topological features, this study seeks to identify trends linking physical graph structure to algorithmic behaviour [16]. Recent work shows that shallow QAOA can exhibit severe reliability and sampling suppression on constrained families unless the circuit is aligned with constraint structure via appropriate encodings and mixing dynamics.[17] Moving beyond idealized benchmarks toward realistic, application-driven evaluation is essential for both theory and practice. Theoretically, understanding these interactions can inform better problem encodings and parameter selection [20]. Practically, these insights are a prerequisite for the meaningful adoption of quantum methods in urban planning, resilient infrastructure, and supply chain management [1]. We study the minimum vertex cover problem defined on graph instances derived from urban road networks. Given a graph $G = (V, E)$, the goal is to find the smallest subset of vertices such that every edge is incident to at least one selected vertex. This problem is NP-hard and naturally maps to a quadratic unconstrained binary optimization (QUBO) formulation, making it well suited for both classical brute-force baselines and QAOA-based quantum optimization.

## 2. Literature Review

| **Authors** | **Year** | **Focus** | **Results** | **Limitations** | **Addressal of this research paper** |
|---|---|---|---|---|---|
| Spatial networks, Barthelemy, M. | 2011 | Establishes the physics and mathematical foundations of networks | Proves that spatial graphs are bounded by physical geography | Does not address the computational complexity or the performance of quantum algorithms when applied to these specific topologies | Shows that the "spatial constraints" actually create unique "energy landscapes" in QAOA |
| Quantum computing for resilient energy systems and infrastructure. Ajagekar, A., & You, F. | 2019 | Investigates the potential of quantum algorithms to optimize complex infrastructure systems | Quantum computing could theoretically solve large-scale combinatorial problems in infrastructure more efficiently than classical methods | Focuses on high-level infrastructure applications without deeply investigating how the specific local geometry of the network itself creates bottlenecks for the algorithm at low circuit depths | Shows that specific *topological DNA* of the network dictates whether the quantum solution will be reliable or fail. |
| Reachability Deficits in Quantum Approximate Optimization. Akshay et al | 2020 | Investigates the fundamental limitations of low-depth QAOA in reaching the optimal solution for certain graph instances | Proves that for certain dense or highly constrained graphs, the QAOA ansatz at p=1cannot physically represent the ground state | Identifies that some graphs fail but does not correlate these failures with real-world infrastructure geometry | Demonstrates that the reachability deficits observed in organic networks (like Lyari) are driven by specific structural metrics |
| Bilateral Spatial Reasoning about Street Networks. Reinhard Moratz | 2025 | Uses network science to perform spatial reasoning on street maps | Shows that urban "sub-regions" can be mathematically classified by their connectivity patterns, | Confined to classical GIS (Geographic Information Systems) | Benchmarks these spatial reasoning categories for quantum algorithms |

| QAOA for Max-Cut requires hundreds of qubits for speed-up. Guerreschi & Smelyanskiy | 2019 | Benchmarks the scalability of QAOA on small vs. large graphs to ne whether a quantum computer might outperform a classical one | Concludes that for small graphs (under 100 qubits), classical brute-force or heuristics are often superior | Overlooks the structure of the problem | Highlights that structure, not just size, is a critical variable in the quantum-classical comparison |
|---|---|---|---|---|---|
| Spatial Quantum Computation in Transportation. Nourbakhshrezaei, Amirhossein | 2023 | Explores using QAOA and other variational algorithms to solve transportation-related graph problems | Embedding of the problem onto quantum hardware is often hampered by the physical layout of the network | Focuses on the application (transportation) rather than the underlying reason for algorithmic failure | Provides a deeper theoretical layer by isolating the urban design as the independent variable |
| The effect of classical optimizers and Ansatz depth. Aidan Pellow-Jarman | 2024 | Compares different classical optimization loops and circuit depths influence on the success of QAOA for Minimum Vertex Cover | Determined that higher depth generally improves accuracy | Uses standardized random graphs, which do not possess the "spatial constraints" found in real city street maps | Substitutes these abstract random graphs with real-world urban subgraphs |
| Performance and limitations of QAOA on large sparse hypergraphs. Joao Basso, David Gamarnik | 2022 | Analyzes the performance of QAOA on sparse graphs | Proved that for sparse, regular graphs, QAOA performs predictably well, but performance degrades as "irregularity" or hyper-edges are introduced | The "irregularity" studied is purely mathematical and lacks the "spatial logic" of real-world urban growth | Shows that the most challenging "irregular" graphs for QAOA are those that result from organic, unplanned city growth |

The existing literature on quantum optimization focuses largely on the theoretical underpinnings of QAOA, such as approximation guarantees and parameter landscapes, typically utilizing idealized random or regular graphs. While foundational works in spatial network science by Barthelemy (2011) define the unique "fingerprints" of geographically constrained networks, and researchers like Ajagekar and You (2019) propose quantum applications for resilient infrastructure, a significant gap remains. Current studies often rely on abstract models that fail to capture the "topological DNA" of real-world urban growth. Theoretical limitations, such as the "reachability deficits" identified by Akshay et al. and the scalability hurdles noted by Guerreschi and Smelyanskiy, suggest that QAOA performance is highly sensitive to graph structure, yet these insights are rarely applied to the structured, non-random geometry of actual street networks.

This synthesis reveals a lack of consensus on how specific spatial features, such as the high degree variance in organic cities versus the regularity of planned grids, influence the optimization landscape. While Pellow-Jarman and Basso highlight how circuit depth and sparsity affect algorithmic success, they overlook the "spatial reasoning" complexities discussed by Moratz. This paper addresses these limitations by integrating comprehensive graph-theoretic metrics with a unified evaluation framework. By moving beyond idealized simulators to analyze geographically grounded instances from Islamabad and Lyari, this study directly connects topological heterogeneity to QAOA reliability, providing a grounded understanding of quantum potential in real-world urban optimization.

# 3. Methodology

## 3.1 Region Selection

Two cities were selected to represent contrasting urban design paradigms. Islamabad serves as an example of a planned city, as depicted in Figure 2 below, characterized by grid-like structure, relatively uniform node degree, and high global organization. Lyari, an organically grown area within Karachi, represents an unplanned topology, as shown in figure 1 below, with irregular connectivity, heterogeneous degree distribution, and localized bottlenecks. These contrasting structural properties allow us to isolate the impact of urban form on algorithmic performance.

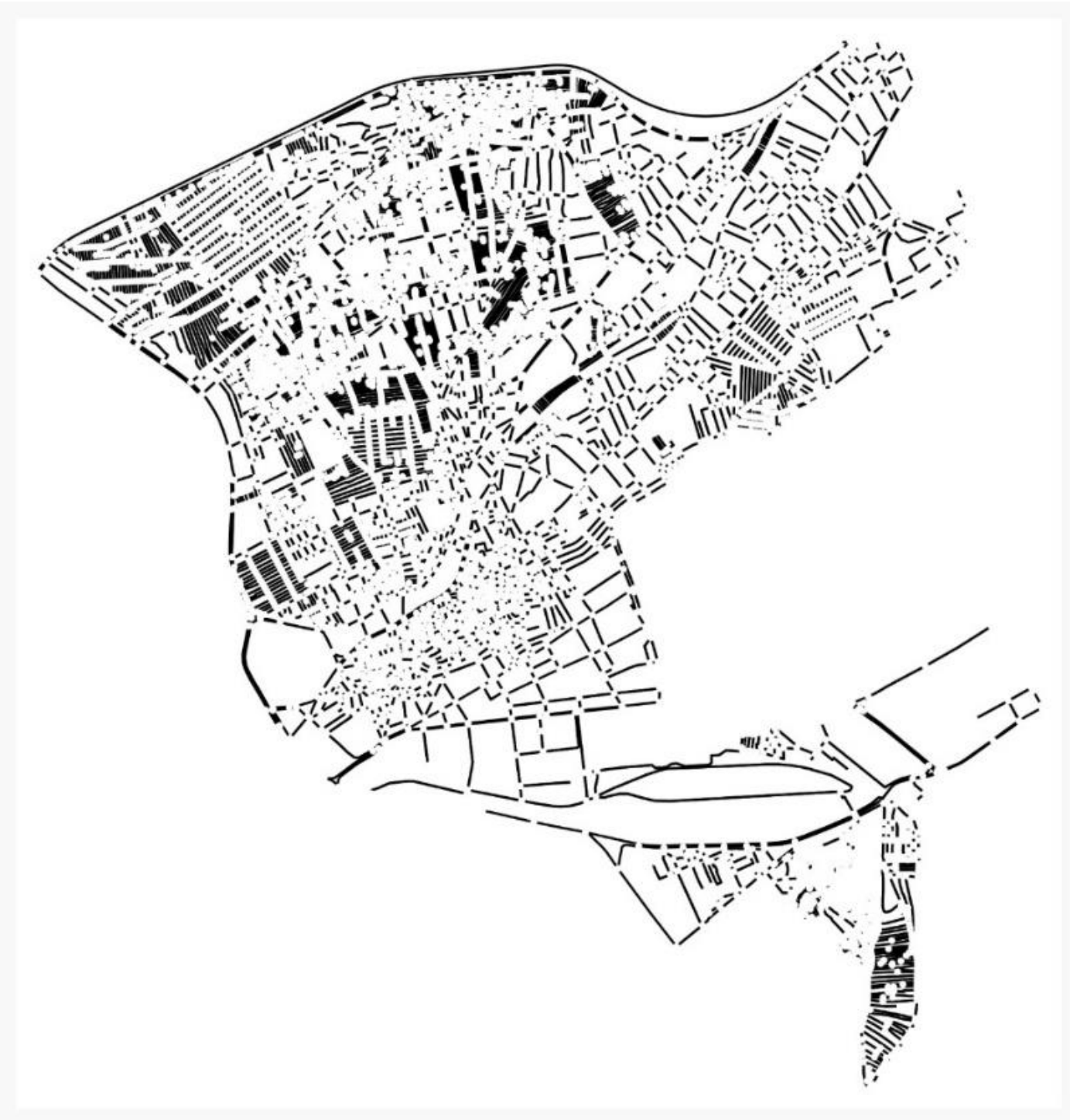

**Figure 1: Map of Lyari, Karachi**

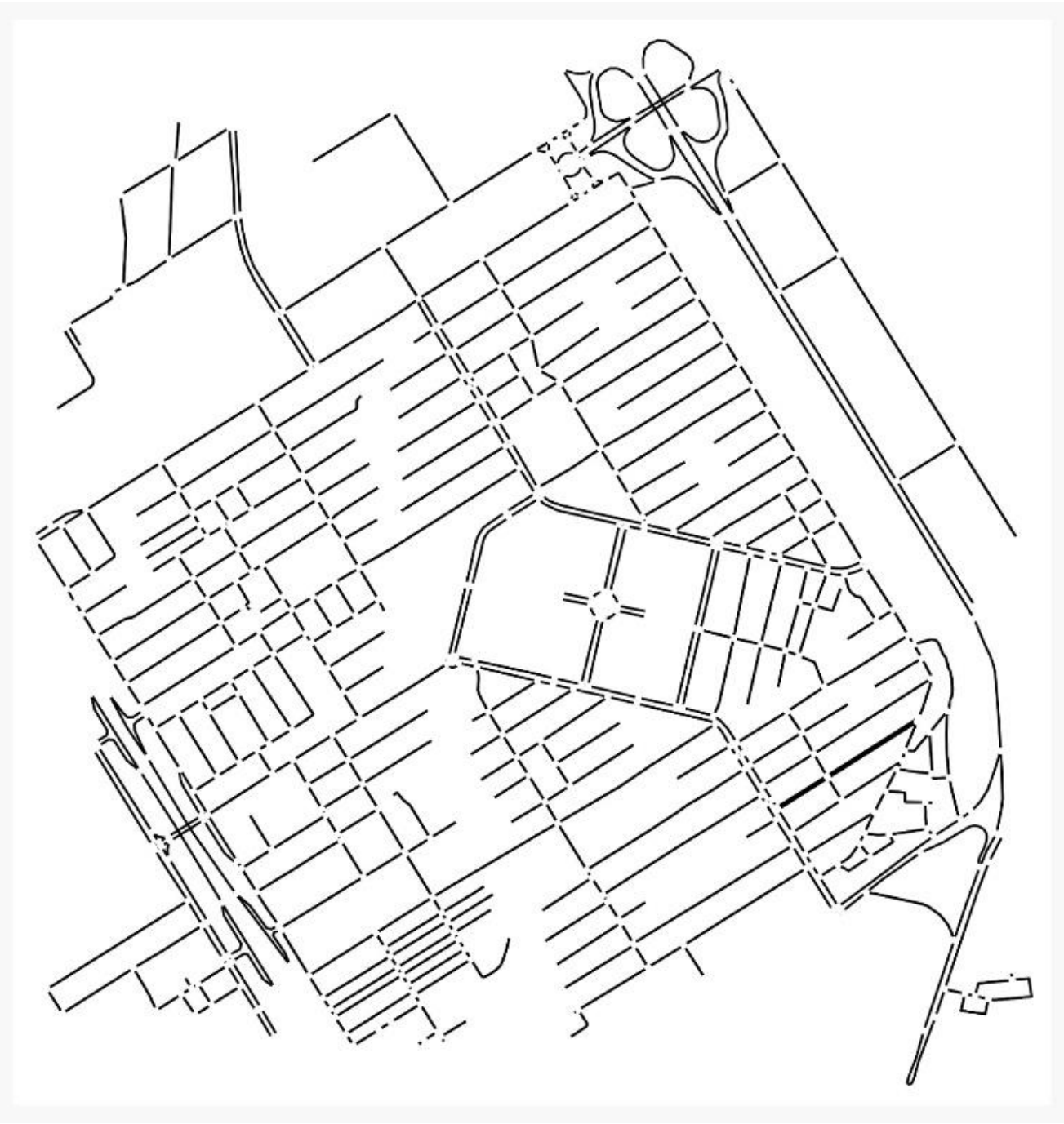

**Figure 2: Map of F-7, Islamabad**

## 3.2 Subgraph Selection and Graph Properties

From each city's street network, we extracted multiple connected subgraphs using a breadth-first search (BFS) procedure. Subgraphs of fixed size N=8 were used for the main analysis, with an extended study at N=10 to evaluate scaling effects. For each subgraph, we computed key structural metrics including degree variance and algebraic connectivity (Fiedler value). These metrics serve as proxies for local heterogeneity and global connectedness, respectively.

```
start_node = random.choice(list(G_full.nodes()))
for N in sizes:
    # 2. UPDATED BFS Logic (Unlimited depth)
    try:
        sub_nodes = list(nx.bfs_tree(G_full, source=start_node).nodes())[:N]
    except:
       sub_nodes = list(G_full.nodes())[:N]

    if len(sub_nodes) < N:
       continue
    G_sub = G_full.subgraph(sub_nodes).copy()
    nodes_order = list(G_sub.nodes())
```

**Code 1: Illustrates BFS subgraph extraction**

```
try:
    fiedler = nx.algebraic_connectivity(G)
except:
    fiedler = 0.0
```

**Code 2: Displays output of degree variance and Fiedler value**

## 3.3 Classical Baseline: Brute-Force Optimization

For each subgraph instance, we computed the exact minimum vertex cover using a brute-force search over all possible vertex subsets. This provided the true optimal solution and corresponding minimum cost. These exact results serve as the reference for evaluating QAOA approximation ratios and for classifying solutions as optimal, suboptimal, or trivial.

```
def solve_classical_mvc(G):
  n = len(G)
  nodes = list(G.nodes())
  for k in range(n + 1):
    for subset in combinations(nodes, k):
      if is_vertex_cover(G, set(subset)):
        return k
  return n
```

**Code 3: Demonstrates exact minimum vertex cover using a brute-force search.**

```
results.append({
      "city": city_name,
      "N": N,
      "topology": topo,
      "true_opt": true_mvc_size,
      "most_probable": {
        "bitstring": most_prob_bs,
        "cover_size": most_prob_size
      },
      "min_energy": {
        "bitstring": min_energy_bs,
        "cover_size": min_energy_size,
        "qubo_energy": qubo_min_energy
      },
      "hamiltonian_expectation": ham_exp,
      "duration": duration })
```

**Code 4: Shows QAOA bitstrings and energies.**

## 3.4 Quantum Approximate Optimization Algorithm

### 3.4.1 QUBO Formulation

The minimum vertex cover problem was encoded as a QUBO objective function by associating a binary variable $x_i$ with each vertex. Penalty terms were added to enforce edge coverage constraints, while linear terms minimized the number of selected vertices. The resulting cost function takes the standard quadratic form and is suitable for direct mapping to a quantum Hamiltonian.

### 3.4.2 Ising Hamiltonian

The QUBO formulation was transformed into an Ising Hamiltonian using the standard change of variables defined as,

$$x_i = \frac{(1 - z_i)}{2} \quad \text{Eq. 1}$$

where $z_i \in \{-1, +1\}$. This yielded a problem Hamiltonian $H_C$ expressed as a sum of single-qubit $Z$ terms and two-qubit $ZZ$ interactions. A transverse-field mixer Hamiltonian $H_M = \sum_i X_i$ was used to drive transitions between computational basis states.

### 3.4.3 QAOA Circuit Construction

We implemented QAOA at depth $p = 1$. The algorithm begins with a uniform superposition over all bitstrings, followed by one application each of the problem and mixer unitaries parameterized by angles $\gamma$ and $\beta$. Parameters were optimized using a classical optimizer to maximize the expected value of the cost Hamiltonian.

```
def run_qaoa_statevector(H, offset, G, node_order=None, p=1):
  if node_order is None:
    node_order = list(G.nodes())
  ansatz = QAOAAnsatz(H, reps=p)
```

**Code 5: QAOA circuit is constructed**

```
def cost_func(params):
    bound_circ = ansatz.assign_parameters(params)
    sv = Statevector(bound_circ)
    return sv.expectation_value(H).real

  init_params = np.random.uniform(-np.pi, np.pi, ansatz.num_parameters)
  res = minimize(cost_func, init_params, method='COBYLA', options={'maxiter': 200})

  final_circ = ansatz.assign_parameters(res.x)
  final_sv = Statevector(final_circ)
```

```
hamiltonian_expectation = final_sv.expectation_value(H).real + offset

# Use exact dict for N=10 to find hidden minimums
probs = final_sv.probabilities_dict()

most_probable_bitstring = max(probs, key=probs.get)

qubo_min_energy = float("inf")
min_energy_bitstring = None
```

**Code 6: Circuit optimization**

### 3.4.4 Sampling and Solution Extraction

For each optimized circuit, we performed 10 independent runs with different random seeds. From each run, we extracted the most probable bitstring from the measurement distribution and evaluated its cost relative to the classical optimum.

## 3.5 Plotting and Interpretation of Results

QAOA outcomes were classified into three categories: optimal solutions matching the classical minimum, suboptimal solutions with higher cost, and trivial solutions corresponding to poor or degenerate covers. Approximation ratios were computed as the ratio of QAOA solution cost to the optimal cost. Results were visualized using bar plots and scatter plots to capture both central tendencies and variability across structural metrics.

### 3.5.1 Probability Concentration Behavior

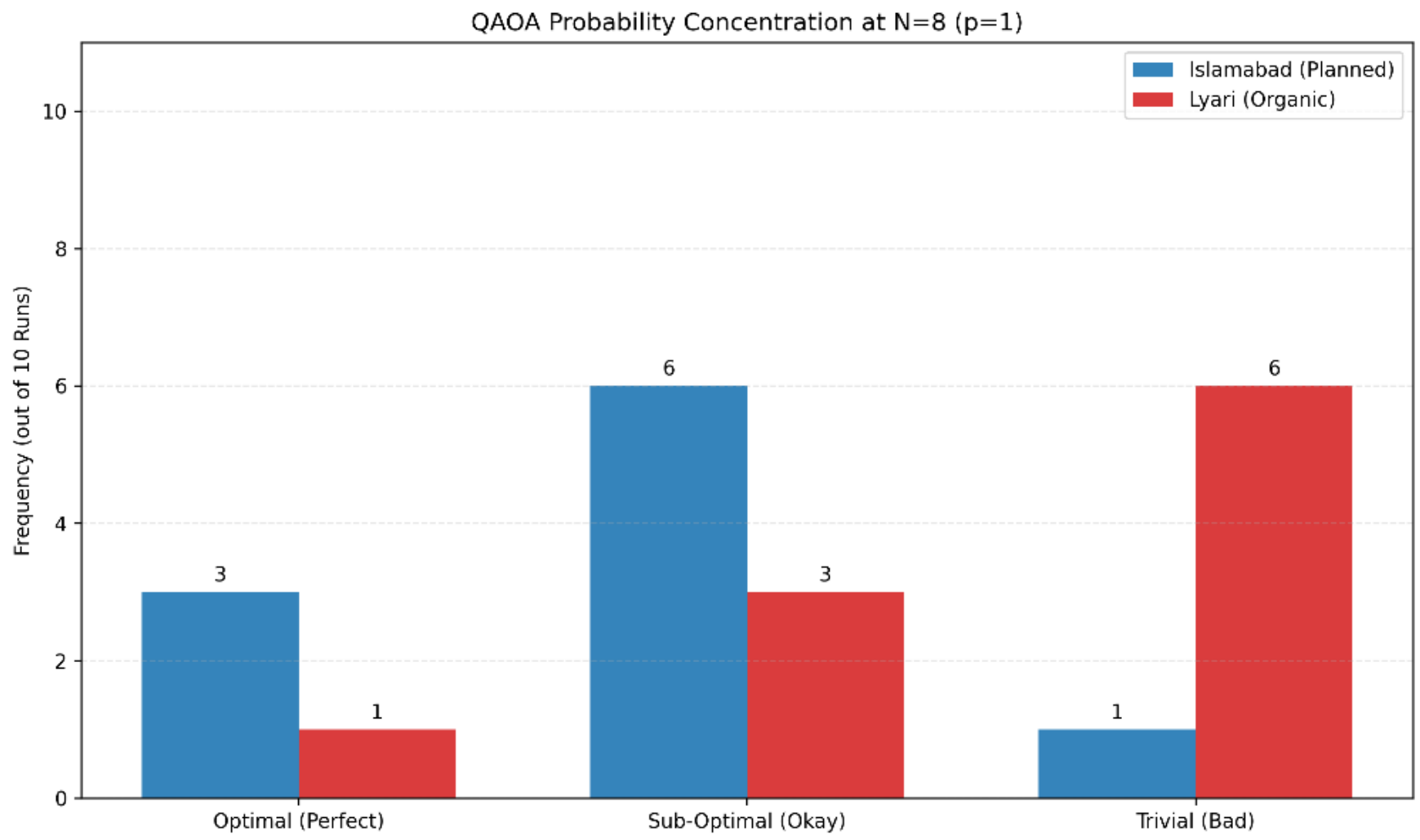

**Figure 1: QAOA Probability Concentration at N=8 (p=1)**

Figure 1 shows the distribution of QAOA outcomes for N=8 subgraphs. The planned topology exhibits a substantially lower frequency of trivial solutions compared to the organic topology. While both cities occasionally yield optimal solutions, the organic network displays a much higher tendency toward poor convergence.

Outcome distribution of shallow-depth QAOA ($p$=1) for BFS-induced subgraphs of size $N$=8. Each bar represents the frequency (out of 10 independent runs with different random seeds) with which the most probable bitstring corresponds to an optimal, suboptimal, or trivial vertex cover. The planned topology (Islamabad) exhibits a significantly lower rate of trivial convergence compared to the organic topology (Lyari).

### 3.5.2 Algebraic Connectivity and Performance Variability

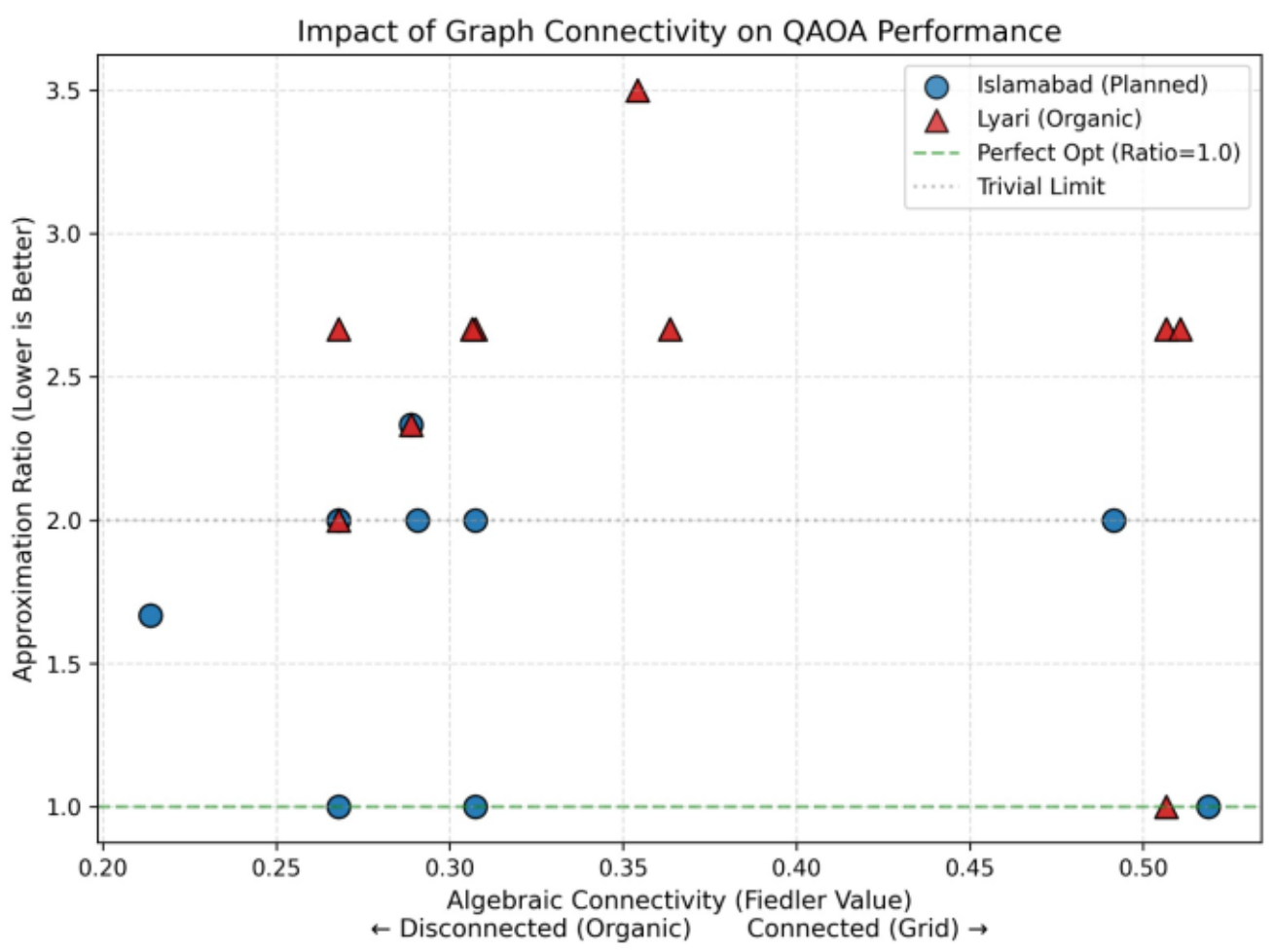

**Figure 2: Impact of graph Connectivity on QAOA Performance**

Figure 2 examines the relationship between algebraic connectivity and approximation quality. No monotonic trend is observed for either city. However, a clear difference emerges in the spread of results. Planned subgraphs remain within a narrow approximation range, while organic subgraphs show high variance and multiple extreme outliers.

Scatter plot of QAOA approximation ratio versus algebraic connectivity (Fiedler value) for BFS-induced subgraphs of size $N$=8. While no monotonic relationship between connectivity and performance is observed, the organic topology exhibits substantially higher variance in approximation quality, including several high-error outliers. This indicates that connectivity alone does not determine QAOA performance. Instead, the organic topology introduces instability that manifests as heteroscedastic behaviour rather than systematic degradation.

### 3.5.3 Degree Variance and Structural Heterogeneity

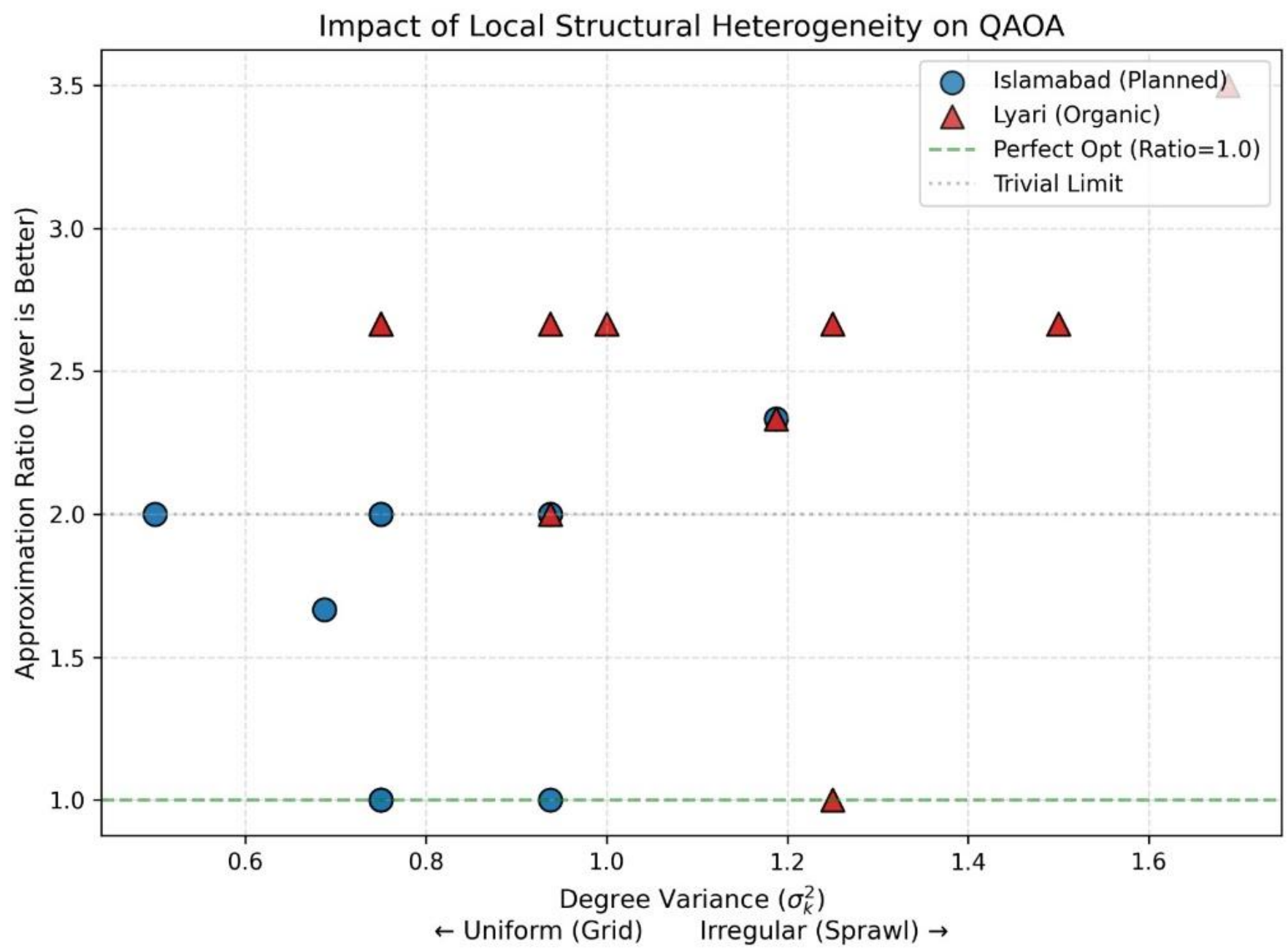

**Figure 3: Impact of Local Structural Heterogeneity on QAOA**

**Figure 3** highlights the role of degree variance. Subgraphs with higher variance, predominantly from the organic city, tend to produce worse approximation ratios. In contrast, planned subgraphs cluster around lower variance values and more stable performance.

Impact of local structural heterogeneity on QAOA performance (N=8, p=1). The X-axis represents the degree variance ($\sigma^2$) of the subgraph.
(Blue Circles) Islamabad: The planned network exhibits low degree variance ($\sigma^2$=0.5) and a constrained performance profile, with approximation ratios consistently bounded below 2.4.
(Red Triangles) Lyari: The organic network exhibits high degree variance ($\sigma^2 > 0.9$). While optimal solutions are occasionally found (r=1.0), this regime is characterized by high algorithmic instability, with frequent divergence to high-energy states (r > 2.5). This suggests that local heterogeneity roughens the optimization landscape, increasing the probability that shallow-depth circuits get trapped in suboptimal local minima.

### 3.5.4 Impact of Local Structural Heterogeneity on QAOA Performance

The relationship between local graph structure and algorithmic stability is further detailed in the fourth result, which analyzes approximation ratios against degree variance ($\sigma^2$).

- **Variance as a Performance Predictor**: Subgraphs with lower degree variance, typical of the planned grid structure in Islamabad, cluster toward more stable approximation ratios.
- **Impact of Urban Sprawl**: In contrast, subgraphs from Lyari exhibit higher degree variance, reflecting an irregular "sprawl" topology that correlates with a wider spread in performance.
- **Tendency Toward Trivial Limits**: The data indicates that as degree variance increases, there is a marked tendency for the algorithm to hit the "Trivial Limit," particularly within organic topologies.

- **Reliability Gap**: While some high-variance subgraphs still reach near-perfect optimization (Ratio = 1.0), the high concentration of poor-performing outliers in irregular networks suggests that structural heterogeneity is a primary driver of optimization instability.

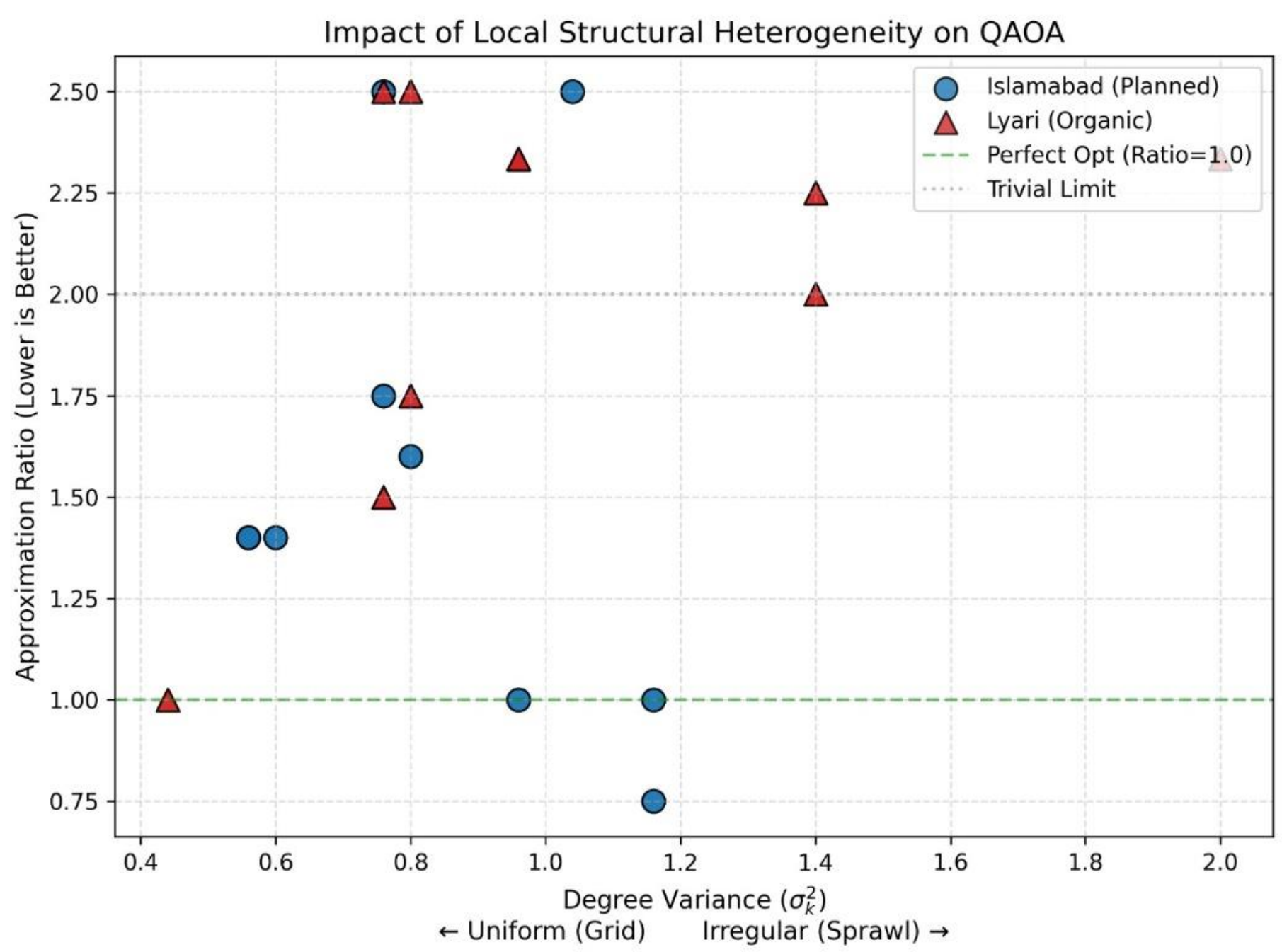

**Figure 4: Impact of Local Structural heterogeneity on QAOA with degree variance at n=10**

Repeating the degree-variance analysis for larger subgraphs (n = 10) reveals a partial attenuation of the clean separation observed at n = 8. While the planned network continues to exhibit more stable performance on average, the degree variance distributions increasingly overlap between the two cities. This suggests that degree variance is a strong but not exclusive predictor of QAOA instability at larger graph sizes, where local structural effects are partially averaged out.

Importantly, the highest approximation ratios (≥ 2.0) remain predominantly associated with the organic topology, indicating that structural heterogeneity continues to act as a risk factor even as problem size increases.

### 3.5.5 Effect of Problem Size

Repeating the degree-variance analysis for N=10 reveals a partial attenuation of the separation observed at smaller sizes. While the planned network remains more stable on average, the distributions increasingly overlap. Importantly, the worst approximation ratios remain concentrated in the organic topology, indicating that structural heterogeneity continues to act as a risk factor even as problem size increases.

# 4. Discussion

The results of this study provide a nuanced understanding of how geographical constraints translate into computational challenges for quantum heuristics. A primary observation is that urban topology does not necessarily cap the theoretical potential for high-quality solutions, but it significantly dictates the reliability of achieving them at low circuit depths.

The divergence between the planned grid of Islamabad and the organic growth of Lyari highlights several critical factors:

- **Reliability and Probability Concentration:** Taken together, Figures 1 and 2 indicate that urban topology primarily affects the reliability rather than the average performance of shallow-depth QAOA. While both planned and organic networks occasionally yield optimal solutions, the organic topology exhibits a markedly higher risk of convergence to trivial or high-energy states.
- **Impact of Local Structural Heterogeneity:** Analysis of the relationship between degree variance ($\sigma^2$) and approximation ratios reveals that increased local heterogeneity acts as a significant barrier to performance. As shown in the data, subgraphs from the organic city (Lyari) frequently occupy regions of high degree variance, which correlates with a greater spread in approximation ratios and a higher incidence of results hitting the trivial limit. This failure mode can be interpreted as a finite-depth sampling concentration issue, where structural heterogeneity roughens the landscape and increases the probability of collapsing into degenerate regions.[18]
- **Landscape Deformation:** The absence of a clear correlation between algebraic connectivity and performance suggests that single scalar metrics are insufficient to characterize the complexity of the QAOA energy landscape. Instead, higher-order structural features—such as degree variability, tree-like substructures, and local bottlenecks—may play a more significant role in shaping probability concentration behaviour at low circuit depth.
- **Scalability and Risk Factors:** Even as problem size increases to N=10, structural heterogeneity continues to act as a risk factor. While the distributions of performance begin to overlap, the most severe outliers and worst approximation ratios remain concentrated in the organic topology.

The research highlights a critical link between geographical layout and the computational efficiency of quantum heuristics, specifically demonstrating that while urban topology doesn't limit the theoretical potential for high-quality solutions, it significantly dictates the reliability of achieving them. In comparing the planned, grid-like structure of Islamabad with the organic, irregular growth of Lyari, a clear divergence emerges. While both systems can theoretically yield optimal results, the organic topology of Lyari presents a much higher risk of "trivial" convergence, where the algorithm fails to find a meaningful solution. This suggests that the predictability of a city's layout acts as a safeguard for the stability of shallow-depth algorithms like QAOA.

A central finding is that local structural heterogeneity, specifically degree variance, serves as a primary barrier to performance. In the irregular subgraphs of Lyari, high degree variance correlates directly with volatile approximation ratios and a higher incidence of algorithmic failure. Interestingly, the study finds that standard scalar metrics, such as algebraic connectivity (the Fiedler value), are insufficient on their own to predict success. Instead, higher-order features like tree-like substructures and local bottlenecks play a more dominant role in deforming the quantum energy landscape and disrupting probability concentration. Even as problem sizes scale to N=10, these structural irregularities continue to act as risk factors, with the most severe performance outliers consistently appearing in organic networks.

# 5. Future work

Building upon these findings, future research should explore the evolution of the performance gap between planned and organic topologies as the algorithm scales to higher circuit depths ($p > 1$). While the current study highlights significant reliability issues at $p = 1$, it remains to be seen whether increasing depth can effectively smooth the "roughened" optimization landscapes created by high-degree variance in organic networks. Furthermore, extending the analysis to much larger graph instances beyond $N = 10$ is necessary to determine if structural risk factors are eventually mitigated by statistical averaging or if they pose permanent bottlenecks for real-world urban optimization. Beyond the Minimum Vertex Cover problem, investigating how urban "topological DNA" affects other combinatorial tasks, such as the Traveling Salesperson Problem or routing optimization, would provide a more holistic view of quantum utility in infrastructure management. Finally, transitioning from noise-free simulators to physical NISQ hardware will be critical to understanding how environmental decoherence interacts with the structural heterogeneity of real-world networks. Such efforts could pave the way for topology-aware parameter initialization strategies that explicitly account for local graph metrics to improve algorithmic stability.

.